# Cooling effect in emissions of $^{103m}$Rh excited by bremsstrahlung

**Y Cheng[1], B Xia[1] and C P Chen[2]**

[1)]Department of engineering Physics, Tsinghua University, 100084, Beijing, China
[2)]Department of Physics, Peking University, 100871, Beijing, China

E-mail: yao@tsinghua.edu.cn

**Abstract.** Nonlinear characteristic emissions of Kα, Kβ and γ with a significant triplet splitting at room temperature are observed from the long-lived nuclear state of $^{103m}$Rh excited by bremsstrahlung irradiation. A pronounced phase-transition-like narrowing of the emission profiles occurs immediately after the sample is cooled down to 77 K. The room temperature profiles reappear again abruptly and almost reversibly as the temperature drifts freely back to approximately the ice point after the filling of liquid nitrogen is stopped. These emission properties at 300 K and at low temperature may indicate that the $^{103m}$Rh nuclei are in collective states.

## 1. Introduction

In the previous reports, interesting anomalies with the characteristic emissions, including Kα, Kβ and γ, at room temperature from the long-lived nuclear state of $^{103m}$Rh excited by bremsstrahlung irradiation have been observed [1-3]. There are several methods to excite the long-lived nuclear states with different species of nuclei, as reported in the literature: by coulomb excitation [4], the (γ,γ′) bremsstrahlung excitation [5-8], the (γ,γ′) photo-activation [9], and the (n,γ) neutron excitation [10]. Although $^{107m}$Ag [5,6,7,9], $^{109m}$Ag [5,6,9], and $^{103m}$Rh [8,9,10] obtained by (γ,γ′) and (n,γ) have been investigated, none of them addressed the issue of anomalous emissions as reported in our previous works [1-3] and the study presented in this paper.

For the inversion density of a two-level model of the $^{103}$Rh isotope excited by the bremsstrahlung irradiation, linear and nonlinear regimes have been identified and reported [1]. By low bremsstrahlung exposure of $^{103m}$Rh, the observed luminosity of the characteristic emissions is linearly proportional to the exposure, denoted as Regime I, whereas in the regime of high exposure, denoted as Regime II, it increases nonlinearly. The luminosity, which is measured directly in the experiment, always decays at room temperature with a time constant just about the natural time constant of $^{103m}$Rh, which is 4857 s. It reflects that the inversion density is a linear function of the luminosity as expected. Therefore, in the nonlinear regime, the inversion density as determined by the measured luminosity increases nonlinearly with the radiation exposure.

A typical gamma-spectrum is given in the supporting online martial. According to the analysis of spectral deformations, stationary splittings show up for all of the characteristic emissions, including Kα, Kβ and γ, at room temperature with the same feature [1]. In the linear regime, the γ splitting energy of triplet model remains at the same relatively small value, around 50 eV [1], while in the highly nonlinear regime, it significantly opens up, over 400 eV, as will be demonstrated and analyzed in detail in the present work. In addition, the enhancement of the splitting energy is irreversible even when the inversion density of $^{103m}$Rh decays down to the level in the linear regime within the time scale of the measurement, ~ 3 hours. Since the separation of splitting peak 50 ~ 200 eV, is smaller than the full width half maximum (FWHM) of the detector, ~ 400 eV, we are not able to fully resolve the triplet splitting for the four measurements presented in the previous work [1]. Two models, the triplet (T model) and the doublet (D Model) splittings, are thus applied to demonstrate the tendency in the enhancement of energy splitting in the nonlinear regime. With a further enlarged splitting energy for the data in the even more highly nonlinear regime in this work, the T model is found to manifest itself. This important observation reveals that the triplet splittings are related to the long-lived nuclear state of $^{103m}$Rh. It indicates that the enhancement in the triplet splitting energy is strongly correlated to the nonlinear pumping efficiency and that any other contribution with different time constants, such as the contributions from the pile-up, the line noise or the radioactive impurities, is unlikely to be responsible for the triplet splitting of the characteristic emissions of $^{103m}$Rh.

In addition to the measurement at 300 K, the characteristic emissions of $^{103m}$Rh by a high rate of exposure to bremsstrahlung irradiation, which is in the highly nonlinear regime, are also investigated at low temperature, from 77 to 300 K, by cooling using liquid nitrogen (LN$_2$). The time-resolved gamma spectroscopy is taken by a second detector [11], denoted as Detector A, different from the one used in Ref. [1], denoted as Detector B. The measurements taken by Detector B are only performed at room temperature, while the measurements taken by Detector A in the present work are carried out at 77 K and in the subsequent warming-up period to room temperature. The results of measurements at 300 K by these two different detectors are consistent with each other in showing the energy splittings of the emission profiles by the analysis of spectral deformation. For further intercomparison, the detection efficiencies of these two detectors have been normalized by the same radioactive source of $^{109}$Cd as a baseline.

In the present work, a couple of interesting properties with the characteristic emissions of Kα, Kβ and γ are reported in the highly nonlinear regime, both at room temperature and at low temperature from 77 to 300 K. At room temperature, with merely 3% of incremental bremsstrahlung exposure, the Kα luminosity is doubled in comparison with that of the data point by the maximum bremsstrahlung exposure reported in the previous work [1]. In the mean time, the corresponding stationary splitting energy analyzed in time evolution is found with even more pronounced nonlinearity, which has been mentioned briefly in the supporting online material of the previous report. In the extreme limit, data points in the highly nonlinear regime exhibit the γ splitting energies exceeding 400 eV, which is resolvable directly by the measurement of the detector without resorting to the data analysis. The cooling effect on the spectral profiles of $^{103m}$Rh emissions is studied by changing the sample temperature. We also study the characteristic emissions of $^{195m}$Pt, which are the Pt impurity with the concentration below 5 ppm in our Rh sample. The inversion density of $^{195m}$Pt excited together with $^{103m}$Rh by the bremsstrahlung also increases in Regime II.

## 2. Experiments

The square rhodium sample with dimensions of 2.5×2.5×1 mm$^3$ is irradiated by the bremsstrahlung from a 6-MeV linac [1-3]. The sample is polycrystalline with 99.9% purity (Goodfellow Rh00300). The natural abundance of $^{103}$Rh is 100% so that almost all of the nuclei in crystal are identical. The Pt impurity concentration specified by the vendor is 5 ppm or less. The procedure of experiment is the same as in Ref. [1], except for a slight modification to adapt the cooling set up by LN$_2$. For the measurements at low temperature, the sample is 2-mm away from the beryllium window of the detector head, which sticks into a plastic container holding $LN_2$, as shown in figure 1. The separation space of 2 mm is accounted for in the calibration for the detecting efficiency of the detector described in the next paragraph. The bremsstrahlung intensity is fine-tuning by varying the duration of the electron-beam (e-beam) macropulses and its repetition frequency with the e-beam energy fixed at 6 MeV. There is no emissions of $^{104m}$Rh detected, indicating that the endpoint energy of the e-beam is below 6.2 MeV, as discussed in Ref. [1]. The exposure rate is monitored by the dosemeter (PTW UNIDOS) with a Farmer ionization chamber (PTW TW30013) located 1 m behind the sample. The same dosemeter, as used in the previous work [1], is calibrated against a standard source every year. The irradiation spot is Gaussian with a FWHM of 8 mm in diameter at the sample center, as depicted in figure 1. The irradiation is done for 2 h, immediately followed by the data acquisition for a period of 3 h. Each full spectrum taken for 3 hours is divided into 180 sub-spectra with a data-taking time of 1 min each. The time resolved behavior of the characteristic emission is thus revealed by these 180 sub-spectra. In the measurement, the detector is horizontally leveled and oriented in the north-south direction, roughly parallel to the earth magnetic field, as shown in figure 1. Detector A is a low-energy high-purity germanium (HPGe) detector (CANBERRA GL0210P) with a 200-mm$^2$ active area and an optic feedback pre-amplifier (CANBERRA 2008 BSL) which are covered by the beryllium window. The data acquisition system consists of a linear amplifier (ORTEC 572A) and a multichannel analyzer (MCA, ORTEC 917A). The energy spectra are taken by a channel width of 43.5 eV, which is different from that of 25.7 eV for the previous report [1]. For a comparison on the spectra in both experiments, we have rescaled the amplitude of normalized profiles by a factor of 25.7/43.5 per channel.

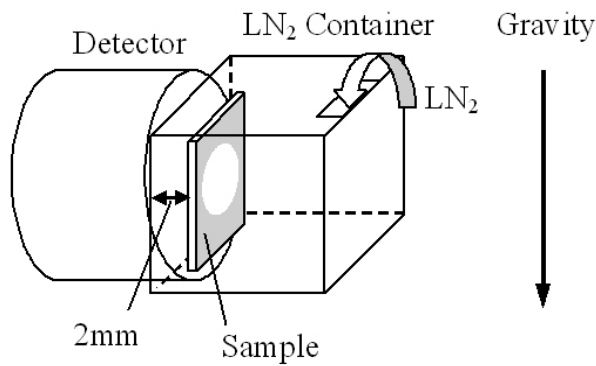

Figure 1. Schematic diagram for the configuration of measurement. The central white spot on the sample stands for the irradiation location. The liquid nitrogen ($LN_2$) is filled from outside into the plastic container. The separation is 2 mm between the sample and the beryllium window of the detector. The relative orientation of the set-up with respect to the gravity is indicated by the arrow.

We have calibrated the detecting efficiencies for Detector A used in this work and Detector B in the previous work by the same $^{109}$Cd source, as shown in figure 2. The half life of $^{109}$Cd is 462.6 days. Its nuclei transmute into $^{109m}$Ag by electron capture to emit the radiation needed. The count rates shown in figure 2 are for the $^{109m}$Ag K$\alpha$ lines at 22 keV, which are in the energy range of interest for our study on the emissions of $^{103m}$Rh. The conversion ratio is obtained from the two calibration lines in figure 2 as 3.834±0.001. In addition, the long term detection stability of Detectors A and B is determined as about $10^{-4}$ and $10^{-5}$, respectively, by the variation of peak position with the fixed energy of $^{109}$Cd source by several calibration measurements performed within a couple of days in a temperature-controlled environment. The stability for Detector A within the measurement period of 3 hours, however, is also on the level of 1 eV (~$10^{-5}$). The background count rate inside the lead shielding of 10-cm thickness is very low in the energy range of interest (0~40keV), about 0.1 cps, which is the same as reported in the previous experiments [1]. During the cooling period, the background count rate is doubled due to the small opening for the filling of LN$_2$ [11]. For a typical cooling experiment, the energy spectrum of $^{103m}$Rh is taken at room temperature for the initial thirty minutes right after the irradiation excitation. Then, the sample is cooled down to 77 K by constantly filling the LN$_2$ into the plastic container (figure 1) for 1 h. With this configuration, the sample is immersed in the LN$_2$ throughout the one hour period of LN$_2$-filling and hence it is cooled from both sides. By rough estimation, the heating power is about 1μW due to the 40-keV nuclear transition, assuming the inversion density of $10^{12}$ cm$^{-3}$ at the spot center with a decay time constant of 4857 s. With this heating power, we expect a finite temperature gradient established inside the sample. During the cooling, the sample is maintained at 77 K by constantly filling LN$_2$ for one hour. After the LN$_2$ is stopped filling, the sample temperature warms up to room temperature with a relaxation constant of about 15 munities [11].

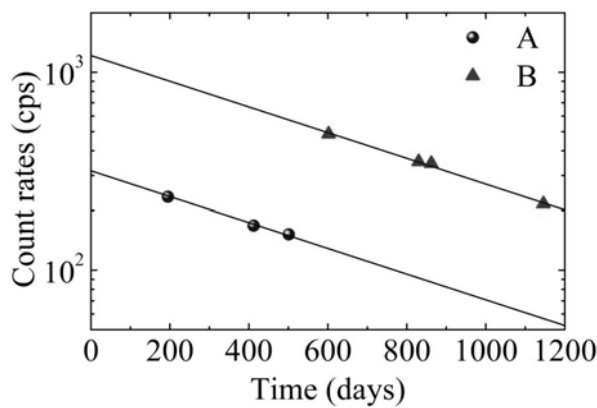

Figure 2. Calibration using $^{109}$Cd source between the two detectors used in the present work (Detector A, filled circle) and in the previous work (Detector B, filled triangle) [1]. The detecting efficiency differs by a factor of 3.834±0.001 at 22 keV.

## 3. Analysis

Figure 3 illustrates the Kα luminosity of $^{103m}$Rh at room temperature as a function of the exposure rate. The luminosity is obtained by extrapolation to time zero corresponding to the end of irradiation calculated from the time-evolution decay behavior of the count rate. For comparison, four data points from previous study [1], as shown by the filled circles, are also plotted. These four measurements are in the low exposure regime. In particular, the first three are in the linear regime and the fourth one shows sign of getting into the nonlinear regime. The two data points analyzed in this report show a highly nonlinear behavior. By increasing the exposure rate only about 3 %, the measured luminosity is doubled, as shown by the open circles in figure 3. Due to the large standard deviation in abscissa, which is estimated as 1.6%, the two data appear almost at the same location in the figure.

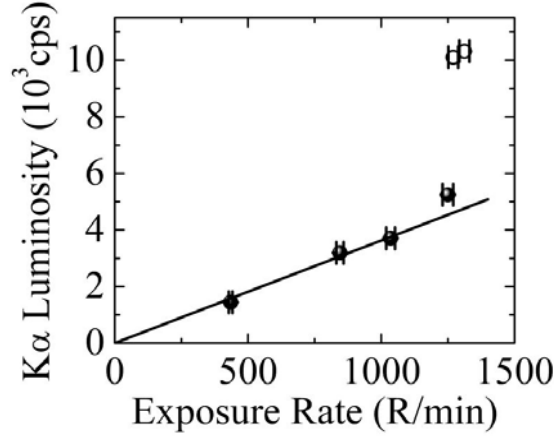

Figure 3. Kα luminosity of $^{103m}$Rh in variation with exposure rate. The Kα luminosity is for the counts in the first minute of measurement at the end of irradiation. It is obtained by extrapolation to time zero from the time-evolution decay behavior of the count rate. The open circles are for the data taken by Detector A in the present experiment, and the filled circles are for the data points taken by Detector B reported previously [1]. The solid line is a linear fit to the first three data points below 1200 Röntgen per minute (R/min).

By adapting the analysis method in the previous report [1], the spectral deformations for the profiles of the three characteristic emissions taken in the present experiment are obtained. The normalized time-dependent spectral deformations, which reveal the deviation of the measured spectral profiles from the calculated normal profiles for the minute-by-minute sub-spectra at the moment, $t_m$, is formulated as,

$$\Delta S_i(E, t_m) = S_i(E, t_m) - \overline{S}_i(E), \qquad (1)$$

in which $i$ stands for Kα, Kβ, and γ, $E$ is the energy of the spectra, $S_i(E, t_m)$ is for the sub-spectra measured for a duration of 1 minute at the moment, $t_m$, $\overline{S}_i(E)$ is for the normal profile with the calibrated FWHM. Both the measured sub-spectra $S_i(E, t_m)$ and the normal profile $\overline{S}_i(E)$ are normalized by $\int S_i(E, t_m) dE = 1$ at the moment, $t_m$, and $\int \overline{S}_i(E) dE = 1$, respectively. The spectral deformations exhibit the property of stationary time evolution. However, after cooling by LN$_2$, the spectral deformations change abruptly at low temperature, switching from one stationary state to another at an unpredictable moment, exhibiting a phase-transition-like behavior. The spectral deformations accumulated for all of the minute-by-minute sub-spectra $S_i(E, t_m)$ within the same stationary state are analyzed by the T model. By this model, the three peaks of splitting for each sub-spectrum at $t_m$ are assumed to have the normal profiles with different amplitudes in (1), formulated as,

$$S_i(E, t_m) = A_{c,i}(t_m)\overline{S}_i(E) + A_{l,i}(t_m)\overline{S}_i(E - \Delta E) + A_{r,i}(t_m)\overline{S}_i(E + \Delta E) \qquad (2)$$

in which the subscripts $c$, $l$, and $r$ stand for the central, the left and the right peaks, respectively. The splitting amplitudes are represented by $A_{c,i}$, $A_{l,i}$, and $A_{r,i}$, respectively. $\Delta E$ is for the energy separation arising from the splitting. To resolve the spectral deformation according to (1), the amplitudes and the splitting energy are obtained by fitting the experimental data using (2) with corresponding normal profiles, $\bar{S}_i(E)$, calibrated for the Detector A.

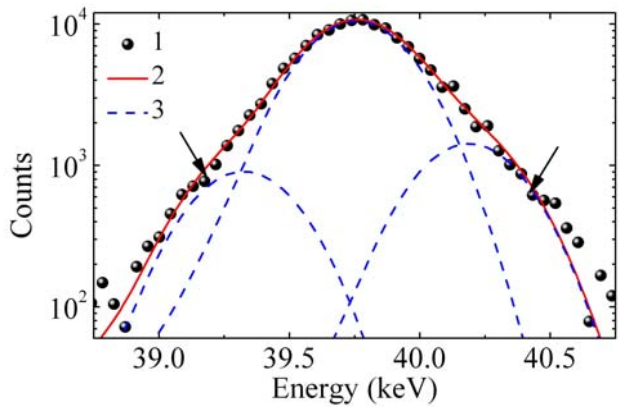

Figure 4. (color online) Triplet splitting of the γ peak at 39.76 keV for the data point with the Kα luminosity of $1.03 \times 10^4$ cps shown in figure 3, Black filled circles are for the measured counts of experimental data (43.5 eV per channel). The solid line in red color is for the fitting result by the triplet splitting model described by (2). The three dashed curves in blue color are for the three individual peaks obtained by the analysis of triplet splitting model with the splitting energy, $\Delta E$. The feature of triplet splitting is obviously revealed by the two kinks as pointed by the arrows at the level of 800 counts. The right shoulder of the γ peak, i.e. the Kα-Kα pile-up located at 40.4 keV, has been removed by the off-line data analysis.

**Table 1.** Splitting energy analyzed by the T model according to (2) for the measurements at room temperature. The Kα luminosity is obtained from the first minute of measurement, as shown in figure 3. The four data points with luminosity < $10^4$ cps is published in the previous report taken by Detector B [1], while the other two data points with luminosity > $10^4$ cps, taken in the present work by Detector A, are fully resolved by the T model. The large error bars of the last two data are attributed to the presence of the gap filled with $LN_2$, as shown in figure 1. These results are plotted in figure 5.

| Kα Luminosity | $\Delta E$ (eV) Kα | $\Delta E$ (eV) Kβ | $\Delta E$ (eV) γ |
|---|---|---|---|
| 1144±3 | 82 | 111 | 46 |
| 3237±4 | 100 | 136 | 50 |
| 3707±6 | 95 | 120 | 59 |
| 5245±2 | 212 | 235 | 165 |
| 10100±100 | 387 | 387 | 430 |
| 10300±100 | 412 | 408 | 430 |

In the previous work, the spectral deformations for the characteristic emissions of the first 4 data points in figure 3 have been analyzed to show broadening effect, without a direct evidence for the triplet splitting [1]. This is because $\Delta E$, which is below 200 eV, is much smaller than the FWHM ~ 400 eV of Detector B. The FWHMs of Detectors A for K$\alpha$ at 20.2 keV, K$\beta$ at 22.7 keV, and $\gamma$ at 39.8 keV are 412, 419, and 463 eV, respectively. For the two data in the highly nonlinear regime with the splitting energy $\Delta E > 400$ eV, the peak profiles show a clear and direct evidence for the triplet splitting, as revealed by the two kinks on both sides of the profiles in figure 4. By the triplet splitting analysis using (2), the free fitting parameters include the three amplitudes, $A_{c,i}$, $A_{l,i}$, and $A_{r,i}$ along with the splitting energy, $\Delta E$. The FWHM is fixed by the calibration of the detector. With the constraint, $A_{c,i} + A_{l,i} + A_{r,i} = 1$, the number of free fitting parameters is three. For the profiles shown in figure 4 and also for another data point in the highly nonlinear regime, the optimum fitting gives $A_{c,i}$ ~ 80%. In figure 4, the solid curve in red color gives a reasonable description for the data points.

The splitting energy analyzed by the T model for all of the data points, including the results for the four measurements reported in the previous paper [1], are listed in table 1. The splitting energy versus the K$\alpha$ luminosity is also plotted in figure 5. For the characteristic emissions in the linear regime (Regime I), the splitting energy stays more or less a constant, about 100 eV for K$\alpha$ and K$\beta$, and 50 eV for $\gamma$. The splitting energy enhances dramatically in the nonlinear regime (Regime II in figure 5), exceeding 400 eV.

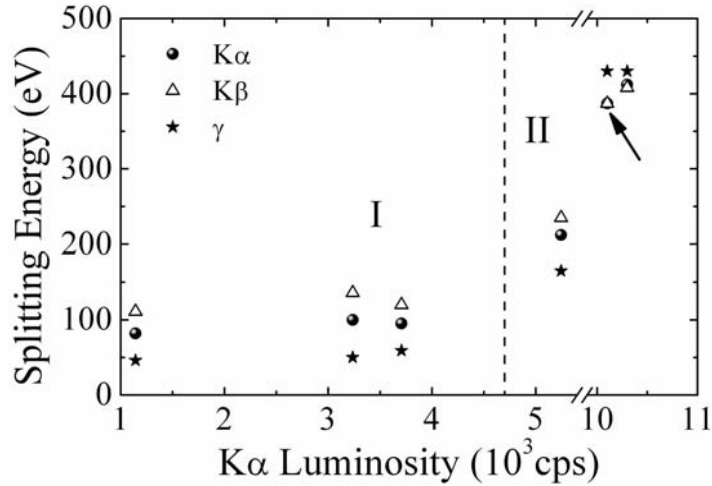

Figure 5. Splitting energy analyzed by T model according to (2) for the three bands K$\alpha$ (filled circles), K$\beta$ (open triangles), and $\gamma$ (filled stars) at room temperature as a function of the observed K$\alpha$ luminosity. The three points on the left hand side are in the linear regime I and the three data points on the right hand side are in the nonlinear regime II. The first four data points on the left hand side are obtained in the previous work [1], while the other two data points on the right hand side are measured in this work. The splitting energies of the three bands are almost coincided for these two data points in the highly nonlinear regime. The data point indicated by an arrow is carried out at 300 and 77 K. The emission properties with temperature variation are presented in figures 6 and 7.

**Table 2.** Splitting energy of the spectral deformations analyzed by T model described by (2) for the three bands corresponding to the five stages presented in figure 7. The uncertainty in $\Delta E$ at stages 1 and 5, which is attributed mainly to the fitting analysis, is on the order of a few eV, whereas in the magnitude of $\Delta E$ at stages 2 to 4 is too small, not resolvable by the detector resolution. The other two results, before and after cooling, stages 1 and 5, show the same level of central splitting amplitude with $A_{c,i} \sim 80\%$, as discussed in the text.

| Stage | 1 | 2 | 3 | 4 | 5 |
|---|---|---|---|---|---|
| Temperature | 300 K | 77 K | 77 K | Warm up | Warm up |
| Time (s) | $0 \sim 1.8 \times 10^3$ | $1.8 \times 10^3 \sim 3.3 \times 10^3$ | $3.3 \times 10^3 \sim 5.5 \times 10^3$ | $5.5 \times 10^3 \sim 8.6 \times 10^3$ | $8.6 \times 10^3 \sim 11 \times 10^3$ |
| $\Delta E(K\alpha)$ | 387 eV | -- | -- | -- | 296 eV |
| $\Delta E(K\beta)$ | 387 eV | -- | -- | -- | 300 eV |
| $\Delta E(\gamma)$ | 430 eV | -- | -- | -- | 318 eV |

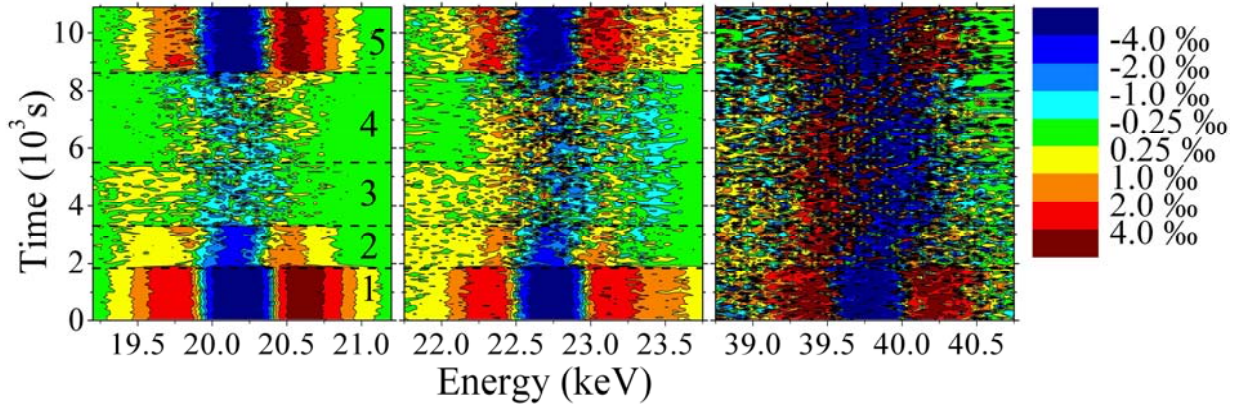

Figure 6. (color online) Time evolution of the spectral deformation for the 180 sub-spectra of three bands, *i.e.* K$\alpha$, K$\beta$, and $\gamma$, analyzed by (1). The evolution is divided into 5 stages separated by the horizontal dashed lines and numbered sequentially in the right edge of the left spectra for the K$\alpha$ deformation. Stage 1 is for the measurement at room temperature. The first-minute room temperature K$\alpha$ luminosity corresponds to the data point indicated by the arrow in figure 4. Stages 2 and 3 are at 77 K, while stage 4 is for the period of stopping filling the LN$_2$, in which the temperature relaxes back to 300 K. The presented amplitude is scaled to 25.7 eV per channel in accordance with that in Ref. [1].

There are four measurements carried out in the highly nonlinear regime with the K$\alpha$ luminosity $\sim 10^4$ cps using the same accelerator parameters for the e-beam. Two of them are shown in figures 3 and 5. The other two data points not shown are measured with the attenuation effect of Cu filters. The measurements with the Cu filters also support the reported behaviors in the highly nonlinear regime at room temperature, *i.e.* K$\alpha$ luminosity $\sim 10^4$ cps, $\gamma$ splitting energy 453±6 eV, and $A_{c,\gamma} = 80\%$ for the data obtained with one Cu sheet, 35 μm in thickness, and K$\alpha$ luminosity $\sim 10^4$ cps, $\gamma$ splitting energy 457±5 eV, and $A_{c,\gamma}$ = 82% for the data obtained with two Cu sheets, both with 35 μm in thickness. We choose not to show these two measurements in figures 3 and 5, since there is no calibration available to precisely determine the Kα luminosity with the attenuation effect by the Cu filters along with the gap filled with $LN_2$. The relationship between the Kα luminosity and the splitting energy is expected to be modified. Interestingly, however, all of the data with the resolved triplet splitting show the same level of central splitting amplitude, $A_{c,i}$ ~ 80%, leading to the argument to set $A_{c,i}$ = 80% by the T model as discussed in Ref. [1].

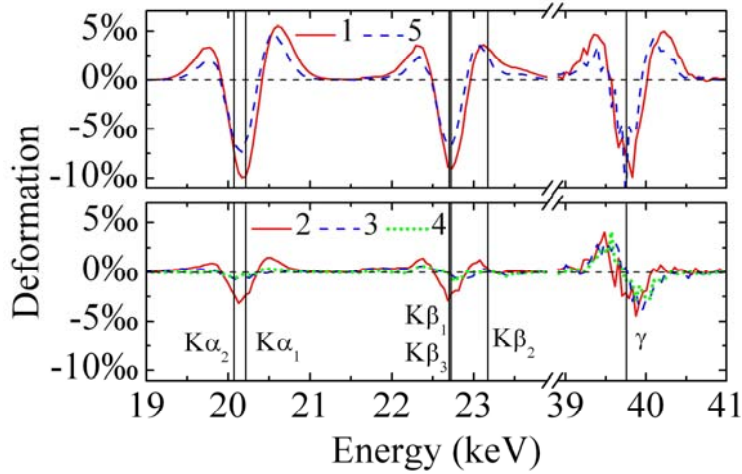

Figure 7. (color online) The spectral deformation calculated by (1) for the five sub-spectra of the three bands accumulated over each data-taking period for stages 1 to 5 shown in figure 6. The deformation spectra for stages 1 (red solid line) and 5 (blue dashed line) shown in the upper plot are more pronounced than those for stages 2 (red solid line), 3 (blue dashed line), and 4 (green dots) at low temperature, shown in the lower plot. The locations of K lines and γ are shown by the vertical lines. Particularly, the $K\beta_1$ and $K\beta_3$ lines almost coincide with each other. The deformations of stages 3 and 4 almost collapse. The presented amplitude is scaled to 25.7 eV per channel in accordance with that presented in Ref. [1].

The following analysis is devoted specifically to the measurement performed with the sample cooled down by $LN_2$. The data point marked by an arrow in figure 5 is the one corresponding to the cooling experiment. Unlike the stationary splitting for the characteristic emissions of $^{103m}$Rh observed at room temperature, the splitting exhibits an interesting phase-transition-like contraction at low temperature. After the bremsstrahlung irradiation, the data-taking lasts as long as 3 hours. The time-evolution property of the three-band emissions is revealed by dividing the spectrum into 180 sub-spectra, each having one minute of data-taking time. According to the properties of the spectra corresponding to different cooling conditions of the $^{103m}$Rh sample, the time-evolution of the emissions is divided into five stages. Figure 6 shows the time evolution behavior before and after the cooling for the energy splitting analyzed by (2) for the three bands of Kα, Kβ, and γ. It consists of each minute-by-minute sub-spectrum. In stage 1 before the cooling, from the time $t$ = 0 to $1.8 \times 10^3$ s (0.5 h), the temperature is at 300 K. The three bands of Kα, Kβ, and γ, exhibit the triplet splitting properties as discussed in the preceding paragraph. After the sample is cooled down to 77 K by adding $LN_2$ into the container to submerge the sample at the instant $t = 1.8 \times 10^3$ s, the spectral deformation changes immediately in response to the sudden change of temperature at the beginning of stage 2. The LN$_2$ level is maintained by constant refilling of LN$_2$ to submerge the sample throughout the entire stages 2 and 3, ensuring that the sample stays at 77 K. The splitting amplitude abruptly reduces for the three bands simultaneously as the sample temperature cools down. Surprisingly, a phase-transition-like, further contraction of K$\alpha$ and K$\beta$ deformation appears even during the period of constant temperature at $T$ = 77 K without any variation of the experimental conditions, as shown in figure 6 at the end of stage 2. Besides the triplet splitting properties by the analysis of spectral deformation, the original profiles for the three characteristic emissions as the direct evidence without the analysis also exhibit a pronounced phase-transition-like behavior at low temperature, see the online supporting material. .

Throughout the low temperature stages, from stages 2 to 4, the γ peak appears to shift to the left side, as if the left part of triplet splitting is asymmetrically enhanced, as shown in figure 7. This abrupt change of symmetry for the γ peak is identifiable due to the symmetrical shape in nature even with deformation [1], whereas the abrupt change of symmetry for K$\alpha$ and K$\beta$ peaks is less identifiable due to their asymmetrical shape in nature. The abrupt transition of splitting symmetry is generally observed by cooling down the sample to 77 K with Detector A and also observed by the enhanced exposure at room temperature with Detector B. It is thus unlikely attributed to the instrumental response of the HPGe detector to the temperature variation. The physics resulting in this variation and the phase-transition-like behavior are yet to be studied further. We will focus on this in the future. Meanwhile, there is an additional asymmetric tail showing up on the left side of the deformation spectra in stages 2 and 3 as shown in figure 6. It is particular pronounced for K$\alpha$ and K$\beta$. The extra tails disappear in stage 4 once the LN$_2$ is stopped filling. It is attributed to the forward part of the Compton scattering by LN$_2$ in the 2-mm gap, causing the asymmetric deformation $> 2 \times 10^{-4}$ per channel displayed with the yellow color online on the lower energy side. The relative intensity of Compton tail for the K$\beta$ is stronger than that for the K$\alpha$, since the cross section of Compton scattering is larger and the photo-electric attenuation is smaller for the K$\beta$ than for the K$\alpha$ [12]. This indicates that the analysis by spectral deformation is, indeed, sensitive enough to reveal the broadening or splitting of the characteristic emission peak. The LN$_2$ is stopped filling at the beginning of stage 4, $t = 5.5 \times 10^3$ s (~1.5 h). The sample then warms up, drifting back to 300 K from 77 K without any temperature regulation. In the warming period by stopping the filling of LN$_2$ for almost one hour, at the beginning of stage 5 at $t = 8.6 \times 10^3$ s, the spectral deformation abruptly recovers its room-temperature shape simultaneously for the three bands. At the end of stage 4, the temperature measured by a platinum thermistor Pt1000 is above the ice point. The asymmetrical splitting of K lines at room temperature is reported previously, while the γ splitting is quite symmetrical [1]. The moments for the beginning and end of each stage and the splitting energy for the three bands in stages 1 and 5 are summarized in table 2.

Figure 7 shows the results of five sub-spectra accumulated within each stage, from stages 1 to 5, corresponding to curves 1 to 5, respectively. Deformations in almost the same magnitude and shape are observed at stages 1 and 5 at 300 K, as shown in the upper plot of figure 7. However, the magnitude of the splitting is slightly reduced at stage 5, see also figure 6 and table 2. This indicates that the change in deformation induced by cooling is nearly reversible. This is rather different from the irreversible behavior of splitting induced purely by pumping of high exposure at $T$ = 300 K observed within the three-hour period for the measurements in the nonlinear regime. Spectral broadening induced by the line noise has been observed for detector A, of which the preamplifier locates outside the detector head. The abrupt contraction in the broadening of the emission peaks at the moment when the sample is cooled down to 77 K reveals that this wide-opened triplet splitting is not attributed to the line noise. Neither the phase-transition-like spectral change nor the spectral shift is observed with the spectra of $^{109}$Cd during cooling or after cooling at the warming period. This gives strong evidence that the phase-transition-like behavior on change in temperature does arise from the physics of $^{103m}$Rh rather than from any side effect of the detector head by cooling.

As discussed in the previous report [1], K and γ are from the atomic and the nuclear emissions, respectively, with very different time constants. The splitting behavior of the same feature for the atomic and the nuclear emissions in the linear and nonlinear regimes (Regimes I and II) may reveal that their deformations are arising from the collective effect of nuclei in crystal with more than one excited nucleus in interaction. The spectral deformations are stationary at 300 K, whereas they exhibit phase-transition-like behavior on change in temperature. This phase-transition-like behavior may also strongly imply the collective excitation effect for the identical nuclei on the lattice positions. Owing to the extremely small line width, the long-lived Mössbauer effect has been predicted only to exist at ultra-low temperature [13,14]. Two observations, the nonlinear broadening at room temperature and the abrupt change in spectral profiles depending on temperature, may provide an unique support for the long-lived Mössbauer effect at 77 and even at 300 K.

## 4. Impurities

Interesting properties of the characteristic emissions of $^{195m}$Pt, which is the impurity in the Rh sample, are also observed by the irradiation excitation. The long-lived state of $^{195m}$Pt with a half life of 4.02 days is accumulated by the successive bremsstrahlung pumping carried out in three days. The characteristic emissions of $^{195m}$Pt are indentified by the K x-rays (K$\alpha_1$, K$\alpha_2$, K$\beta_{1,3}$, K$\beta_2$ at 66.8, 65.1, 75.7, 77.8 keV) and the $^{195m}$Pt γ-rays (98.9, 129.7 keV). The natural abundance of $^{195}$Pt is 34 % [15]. The pumping efficiency of $^{195m}$Pt is calibrated by a pure Pt sample with the low bremsstrahlung intensity in the linear regime. The efficiency ratio of one $^{103m}$Rh for 0.8 $^{195m}$Pt is obtained by two separate measurements with a pure Rh and a pure Pt samples. The Pt impurity concentration is then determined as 25 ppm, by assuming the same pumping efficiency for the $^{195}$Pt isotope in the Pt sample as well as in the Rh sample. It is more than the value of 5 ppm specified by the vendor. This is, perhaps, attributable to the enhanced pumping efficiency of the $^{195}$Pt impurity embedded in the Rh lattice. Although the impurity concentration of $^{195}$Pt is low, we find a nonlinear increase with the characteristic emissions of $^{195m}$Pt similar to those of $^{103m}$Rh. In the nonlinear regime of this report, the number of $^{103m}$Rh nuclei increases twice with the increasing bremsstrahlung exposure by only 3 %, see figure 3 for the nonlinear increase of K$\alpha$ luminosity with the exposure, while the γ rays of $^{195m}$Pt also increases twice. It indicates that the nonlinear increases of $^{195m}$Pt and of $^{103m}$Rh are by the same factor. Strikingly, however, the intensity of Pt K lines increase by fivefold, which is significantly more than the enhancement of the $^{195m}$Pt γ emissions.

## 5. Conclusion

Nonlinear increase of the $^{103m}$Rh inversion density depending on the bremsstrahlung exposure rate is observed. In the nonlinear regime at room temperature, the splitting of spectral deformation further opens up from the value in the linear regime. Along with the nonlinear increase of the $^{103m}$Rh density, we also observe the increase of impurity density of $^{195m}$Pt, which is excited by the bremsstrahlung irradiation together with $^{103m}$Rh. In addition, abrupt changes for all of the spectral profiles in K$\alpha$, K$\beta$, and γ of $^{103m}$Rh, including the splitting energy and the splitting symmetry, are observed simultaneously by cooling down to 77 K. This phase-transition-like behavior is not only observed at the moment of cooling, but also during the low temperature phase at 77 K, and in the relaxation period of warming up to room temperature. As the temperature warming up close to room temperature, the spectral deformations resume their shape before cooling. Unlike the irreversible opening of splitting at room temperature by the increasing exposure in the nonlinear regime, the phase-transition-like contraction in the splitting induced by the low temperature is quasi-reversible. The phase-transition-like and nonlinear emissions of $^{103m}$Rh nuclei depending on the temperature and the inversion density may indicate the existence of an unusual long-lived Mössbauer effect with the nuclei in the density- and temperature-dependent collective states other than the conventional Mössbauer effect.

**Acknowledgments**
The data is adapted from the thesis of Zhong-Ming Wang. This work is supported by the NSFC grant 10675068.

# Supporting online material

## Cooling effect in emissions of $^{103m}$Rh excited by bremsstrahlung

Y Cheng[1], B Xia[1] and C P Chen[2]

[1)] Department of engineering Physics, Tsinghua University, 100084, Beijing, China
[2)] Department of Physics, Peking University, 100871, Beijing, China

E-mail: yao@tsinghua.edu.cn

The time-evolution spectra, to support the reported phase-transition-like behavior induced by the temperature variation, for the three characteristic emissions of Kα, Kβ and γ from $^{103m}$Rh excited by bremsstrahlung irradiation are presented in figures S1 and S2. The shown data are normalized over the total counts in the peak. The Kα-Kα pile-up (40.4 keV) at the right shoulder of γ peak is removed by the off-line data analysis. A typical gamma-spectrum is illustrated in figure S3.

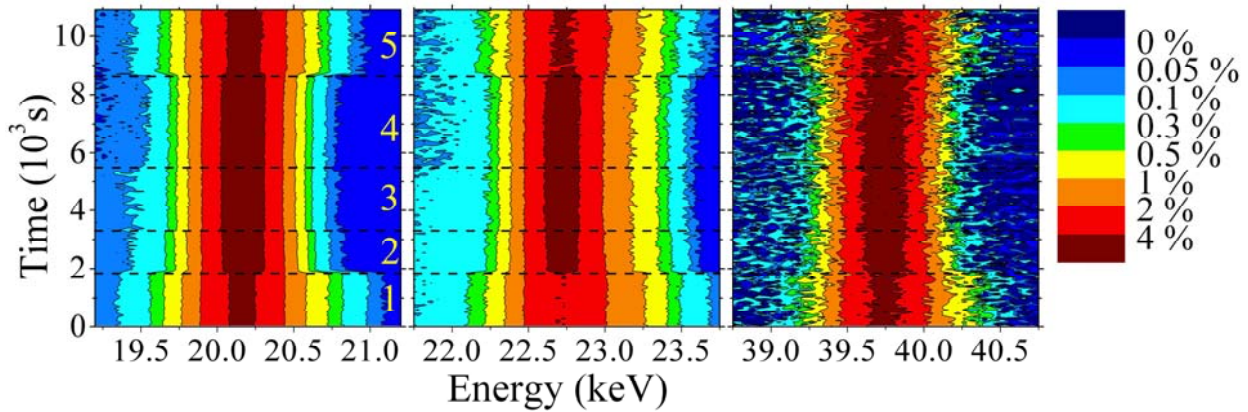

Figure S1: Minute-by-minute sub-spectra of the time-evolution profiles for Kα, Kβ and γ. Each sub-spectrum is normalized over the total counts recorded within each minute. The presented channel width for the data acquisition system is 25.7 eV per channel. Hence, the normalized intensity per channel ranges from a few percent in the center of profiles to about 1 part in a thousand on both sides. The evolution is divided into 5 stages separated by the horizontal dashed lines and denoted sequentially by numbers (white color) in the right edge of the left spectra for the Kα profile. Stage 1 is for the measurement at room temperature. Stages 2 and 3 are at 77 K, while stage 4 is for the period of stopping filling the LN$_2$, in which the temperature relaxes back to 300 K. The Compton tails (~0.1%) on the left hand side of Kα and Kβ peaks disappear at the moment entering stage 4 due to vanishing liquid nitrogen, as discussed in text.

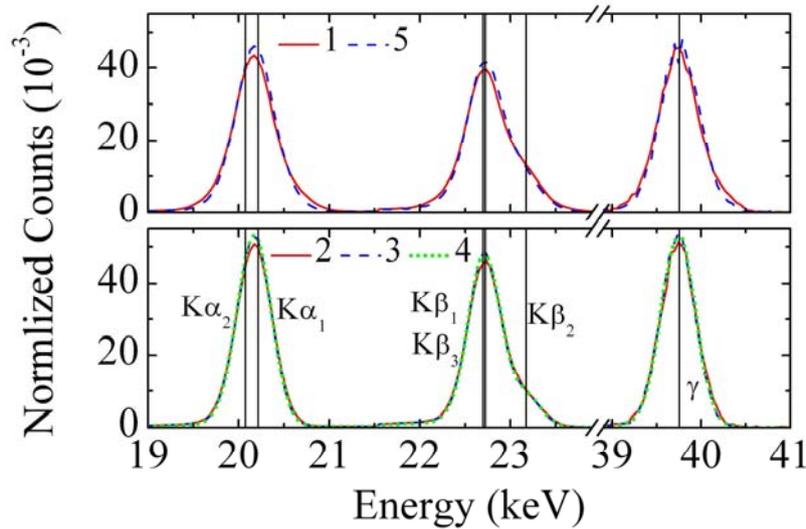

Figure S2: The accumulated profiles for stages 1 to 5 as shown in figure S1 and discussed in the text. For each stage, the profiles are normalized over the total counts within the stage. The locations of K lines and γ are shown by the vertical lines. Particularly, the $K\beta_1$ and $K\beta_3$ lines almost coincide with each other. It is obvious that the peaks become narrow at stage 2, 3, and 4 at low temperature, in comparison to those at stages 1 and 5.

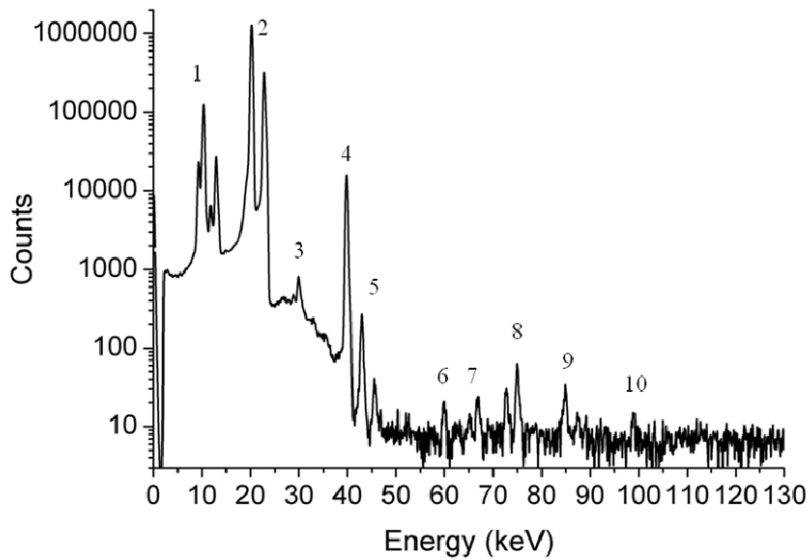

Figure S3: A typical gamma-spectrum of the polycrystalline Rh sample excited by the bremsstrahlung irradiation. Peaks are identified with a numbering in the plot, i.e. 1) escapes of K lines; 2) Kα and Kβ lines; 3) escapes of γ, several peak pile-ups between K lines and their escapes; 4) $^{103m}$Rh γ; 5) peak pile-ups of K lines; 6) peak pile-ups between K lines and $^{103m}$Rh γ; 7) K lines of $^{195m}$Pt; 8) Kα lines of the lead shielding; 9) Kβ lines of the lead shielding; 10) γ of $^{195m}$Pt.